\documentclass[aps,prl,showpacs,groupedaddress,twocolumn,preprintnumbers,amsmath,amssymb]{revtex4}

\usepackage{graphicx}

\begin{document}

\title{Qubit-assisted transduction for a detection of surface acoustic waves near the quantum limit}

\author{Atsushi~Noguchi$^{1,2}$}
\email[]{noguchi@qc.rcast.u-tokyo.ac.jp}
\author{Rekishu~Yamazaki$^{1,2}$}
\author{Yutaka~Tabuchi$^{1}$}
\author{Yasunobu~Nakamura$^{1,3}$}

\affiliation{%
$^{1}$Research Center for Advanced Science and Technology (RCAST), The University of Tokyo, Meguro-ku, Tokyo, 153-8904, Japan,\\
$^{2}$PRESTO, Japan Science and Technology Agency, Kawaguchi-shi, Saitama 332-0012, Japan,\\
$^{3}$Center for Emergent Matter Science (CEMS), RIKEN, Wako-shi, Saitama 351-0198, Japan
}

\date{\today}
             
\begin{abstract}
We demonstrate ultra-sensitive measurement of fluctuations in a surface-acoustic-wave~(SAW) resonator using a hybrid quantum system consisting of the SAW resonator, a microwave (MW) resonator and a superconducting qubit.
The nonlinearity of the driven qubit induces parametric coupling, which up-converts the excitation in the SAW resonator to that in the MW resonator.
Thermal fluctuations of the SAW resonator near the quantum limit are observed in the noise spectroscopy in the MW domain.
\end{abstract}

\maketitle
Hybrid quantum systems have been widely studied in quantum information science~\cite{Kurizkia2015}. 
Coherent control and quantum measurement across heterogeneous subsystems extend the applicability of quantum technologies towards quantum computers~\cite{nakamura2010}, networks~\cite{kimble2008}, repeaters~\cite{briegel1998} and sensors~\cite{degen2017}. 

In solid-state systems, superconducting qubits~\cite{nakamura1999} have been hybridized with a variety of other quantum degrees of freedom.
The qubits provide strong nonlinearity originating in the Josephson effect as an important ingredient in the hybrid systems. 
For example, strong coupling with the systems such as microwave resonators~\cite{walraff2004nat,paik2011}, nanomechanical resonators~\cite{OConnel2010,teufel2015}, bulk acoustic modes~\cite{yale2017}, and paramagnetic~\cite{Zhu2011,kubo2011} and ferromagnetic~\cite{nakamura2015} spin ensembles have been demonstrated. 

Recently, surface acoustic waves (SAW) have attracted much interest as an alternative quantum mode localized on a surface of a material~\cite{delsing2012,leek2016}. 
In piezoelectric materials, SAW can be strongly coupled to electric fields between surface electrodes and are widely applied in compact microwave components because of their small wavelengths and losses~\cite{datta1986}. 
Interaction of a SAW waveguide~\cite{delsing2014} and a SAW resonator~\cite{leek2017} with a superconducting qubit were recently demonstrated. 
SAW can also couple to other physical systems~\cite{lukin2015} such as quantum dots~\cite{santos2009} and NV centers~\cite{wang2016,wang20162} through various form of elastic effects. 
Opto-elastic interaction of SAW opens the possibility to achieve a quantum transducer from microwave photons to optical photons in the telecommunication band~\cite{shumeiko2016,Srinivasan2016,Okada}. 

In this Letter, we report experiments on a hybrid quantum system consisting of a SAW resonator, a superconducting qubit, and a MW resonator. We demonstrate microwave-driven parametric couplings induced by the nonlinearity of the qubit, which serves as a transducer or an interface between the phonons in the SAW resonator and the photons in the MW resonator. The thermal phonons in the sub-GHz SAW resonator are up-converted to the MW frequency range where near-quantum-limited measurement of photons is available. 
We observe thermal fluctuations in the SAW resonator below the mean phonon number of unity with an unprecedented sensitivity. 

\begin{figure}[t]
   \includegraphics[width=8.5cm,angle=0]{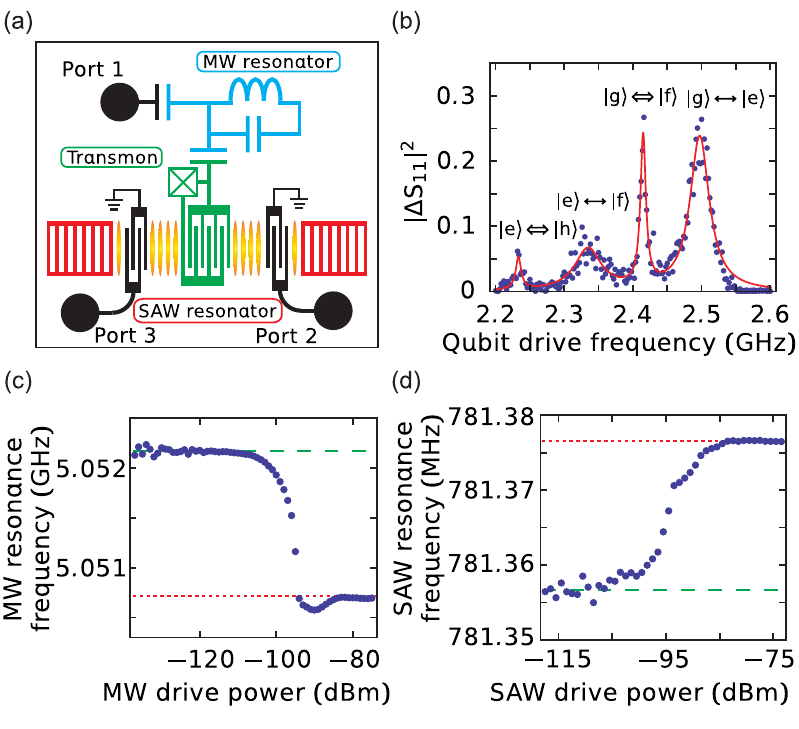}
\caption{Characterization of the SAW-qubit-MW hybrid system. (a)~Schematic of the hybrid system. 
A SAW resonator defined by Bragg mirrors~(red) and a MW resonator (cyan) are coupled to each other via a transmon qubit with an interdigitated transducer (green). The square with a cross represents a Josephson junction.
Port 1 is an external feed line for the MW resonator and ports~2 and 3 are those for the SAW resonator having a spatial mode shown in yellow.
(b)~Qubit spectrum obtained with two-tone spectroscopy. 
The squared change of the reflection coefficient, $|\Delta S_{11}|^2$, of the MW resonator at the resonance is plotted as a function of the qubit drive frequency.
The blue dots represent the data, and the red lines are the Lorentzian fits. 
(c) [(d)] Resonance frequency of the MW (SAW) resonator~(blue dots) as a function of the MW (SAW) drive power. 
In each panel, the red dotted line indicates the resonance frequency of the bare resonator, and the green dashed line indicates that of the dispersively shifted resonator.
}
\label{fig1}
\end{figure}

\begin{figure}[b]
   \includegraphics[width=8.3cm,angle=0]{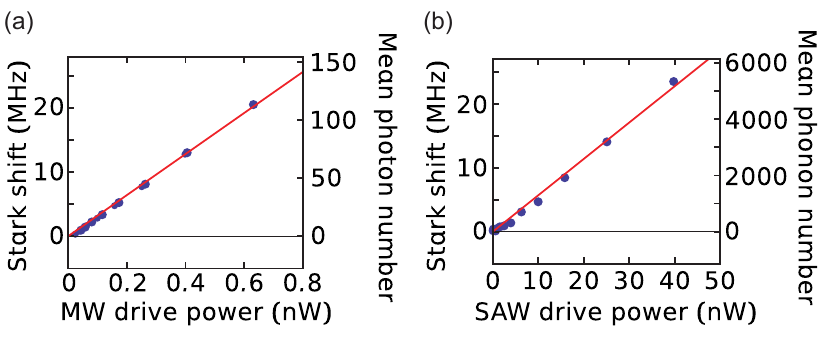}
\caption{Calibration of the photon and phonon numbers in the resonators.
(a)[(b)]~AC Stark shift $\Delta \omega_{q,m}$~($\Delta \omega_{q,s}$) of the qubit as a function of the MW~(SAW) drive power. 
The corresponding mean intra-resonator photon~(phonon) number is indicated on the right axis.
}
\label{fig10}
\end{figure}

The hybrid device schematically illustrated in Fig.~1(a) is fabricated on a quartz substrate and cooled down to 10~mK in a dilution refrigerator (See the details in the Supplementary Information~\cite{supple})).
The Fabry-P\'{e}rot-type SAW resonator, defined by a pair of Bragg mirrors, couples piezoelectrically to a superconducting transmon qubit~\cite{Koch2007} via an interdigitated capacitor. The qubit also couples capacitively to a half-wavelength coplanar-waveguide MW resonator.
The MW resonator has a capacitively-coupled port on the other end (port 1). The SAW resonator has two interdigitated transducer ports (ports 2 and 3) for external coupling.

The hybrid system is modeled with a Hamiltonian under rotating-wave approximation
\begin{eqnarray}
\hat{H}&=&\hbar \omega _m\hat{a}^\dagger\hat{a}+\hbar \omega _e\hat{\sigma}_{ee}+\hbar \omega_f \hat{\sigma}_{ff}+\hbar \omega _s\hat{c}^\dagger\hat{c}\nonumber\\
&+&\hbar g_m (\hat{a}^\dagger \hat{\sigma}_{ge}+\hat{a}\, \hat{\sigma}_{eg})+\hbar g_s (\hat{c}^\dagger \hat{\sigma}_{ge}+\hat{c} \, \hat{\sigma}_{eg})\\
&+&\sqrt{2} \hbar g_m (\hat{a}^\dagger \hat{\sigma}_{ef}+\hat{a} \, \hat{\sigma}_{fe})+\sqrt{2}\hbar g_s (\hat{c}^\dagger \hat{\sigma}_{ef}+\hat{c} \, \hat{\sigma}_{fe})\nonumber
\end{eqnarray}
where $\hbar$ is the Planck constant, $g_m$~($g_s$) is the coupling strength between the MW~(SAW) resonator, and $\omega_e$ and $\omega_f$ are the eigenfrequencies of the first excited state $|e\rangle$ and the second excited state $|f\rangle$ of the qubit.
Here, $\hat{a}$~($\hat{c}$) is the annihilation operator of the MW~(SAW) resonator, and $\hat{\sigma}_{kl} \equiv | k \rangle \langle l | $~($ k,l=g,e,f $) are the operators in the energy eigenbasis of the transmon qubit.

We first characterize the system with spectroscopic measurements~\cite{supple}.
Direct S-parameter spectroscopy is performed on the MW and SAW resonators.
The MW resonator has the resonance frequency of  $\omega _m/2\pi \simeq 5.05$~GHz.
The total linewidth is $\gamma /2\pi = 716$~kHz, and the external coupling rate to port~1 is $\gamma _\mathrm{ex}/2\pi = 152$~kHz.
Similarly, the frequency and the linewidth of the SAW resonator are found to be $\omega _s /2\pi \simeq 781$~MHz and $\Gamma /2\pi = 36.6$~kHz, respectively.
The external coupling $\Gamma_\mathrm{ex}$ through the IDTs connected to ports~2 and 3 are $2\pi \times 60$~Hz each. Thus, the SAW resonator is largely undercoupled. 

Qubit parametrization can be achieved via the spectroscopy of the MW resonator to which the qubit is dispersively coupled.
Figure~1(b) shows a qubit spectrum obtained by the two-tone spectroscopy via the MW resonator~\cite{Schuster2005}.
The eigenfrequencies are $\omega _e/2\pi = 2.53$~GHz and $\omega _f  /2\pi =4.83$~GHz, respectively.
Thus, the qubit anharmonicity $\alpha \equiv \omega_f - 2 \omega_e$ is evaluated to be $ 2\pi \times (-230)$~MHz.
Note that the lines corresponding to the parity-allowed two-photon transitions, $|g\rangle \Leftrightarrow |f\rangle$ and $|e\rangle \Leftrightarrow |h\rangle$,  are also observed, where $|h\rangle$ is the third excited state of the qubit.


Some of the system characterization becomes much easier with the presence of the qubit inside the system.
Coupling between the MW/SAW resonator to the qubit is measured quite conveniently using a simple decoupling technique with a strong drive~\cite{Reed}.
Figures~1(c) and (d) show the drive power dependence of the MW and SAW resonance frequencies, respectively.
A large frequency shift between the low- and high-power limits is observed.
This frequency shift corresponds to $\chi_{i}=g_i^2 /(\omega_e-\omega_i)$~($i=m,s$), which allows us to evaluate the coupling strength between the MW/SAW resonator and the qubit~\cite{Suri}.
From Figs.~1(c) and (d) the coupling strengths $g_{m}$ and $g_{s}$ are evaluated to be $2\pi \times 59$~MHz and $2\pi \times 6.0$~MHz, respectively.

The coupling strength between the SAW resonator and the qubit can also be written as
\begin{equation}
\hbar g_s=eV_{\mathrm{zpf},s}C_\mathrm{IDT}/C ,
\end{equation}
where $e$ is the charge of the electron, $C_\mathrm{IDT} = 67\mathrm{~fF}$ is the capacitance of the IDT coupled to the qubit, $C=84\mathrm{~fF}$ is the total capacitance of the qubit, and $V_{\mathrm{zpf},s}$ is the zero point fluctuation of the surface potential of the SAW resonator.
The capacitances are calculated from the geometrical design of the IDTs~\cite{lukin2015} and are consistent with the single-electron charging energy of the qubit.
From this equation we obtain the zero point fluctuation of the SAW  resonator as $V_{\mathrm{zpf},s}=18\mathrm{~nV}$.
The zero point fluctuation of a SAW resonator is expected to be
\begin{eqnarray}
V_{\mathrm{zpf},s}&=&\phi _0\times \sqrt{\frac{1\mathrm{~\mu m^2}}{A}},
\end{eqnarray}
where $A$ is the mode area of the SAW resonator ($\sim$40,000~$\mathrm{\mu m^2}$). This leads to the surface piezoelectric potential of $\phi _0=3.6\mathrm{~\mu V}$, which agrees well with the theoretical estimate for quartz~\cite{lukin2015}.
From these relations, the coupling strength is written as 
\begin{equation}
g_s/2\pi \approx 0.7\mathrm{~GHz} \times \sqrt{ 1\mathrm{~um^2} / A}.
\end{equation}
Using the mode area of the SAW resonator, the zero point fluctuation in the unit of displacement is estimated as $X_{\mathrm{zpf}}=7\mathrm{~am}$~\cite{lukin2015}.

Dispersive interactions with the MW/SAW resonator makes the qubit subject to the AC Stark shifts proportional to the number of photons and phonons in each resonator. 
We use the effect as a calibration tool for the photon/phonon numbers in the resonators. 
Using the values of $\chi_i$, we estimate the single photon/phonon Stark shifts of the qubit as $\chi_{q,i} = 2 \chi_i \alpha / (\omega_i - \omega_e - \alpha) $ ($i=m,s$)~\cite{Suri} and obtain $\chi_{q,m}/2\pi = 180$~kHz and $\chi_{q,s}/2\pi = 4.4$~kHz.
Figure~2 shows the linear dependences of the AC Stark shifts $\Delta \omega_{q,i}$ on the MW and SAW drive powers.
On the right axes, the corresponding intra-resonator photon/phonon numbers $\langle n _i\rangle = \Delta \omega_{q,i} / \chi_{q,i}$ are indicated.
The values of the Stark shifts per photon/phonon are smaller than the decay rates of the resonators and the qubit. 
Thus, the hybrid system is in the weak dispersive coupling regime. 

\begin{figure*}[bt]
   \includegraphics[width=16cm,angle=0]{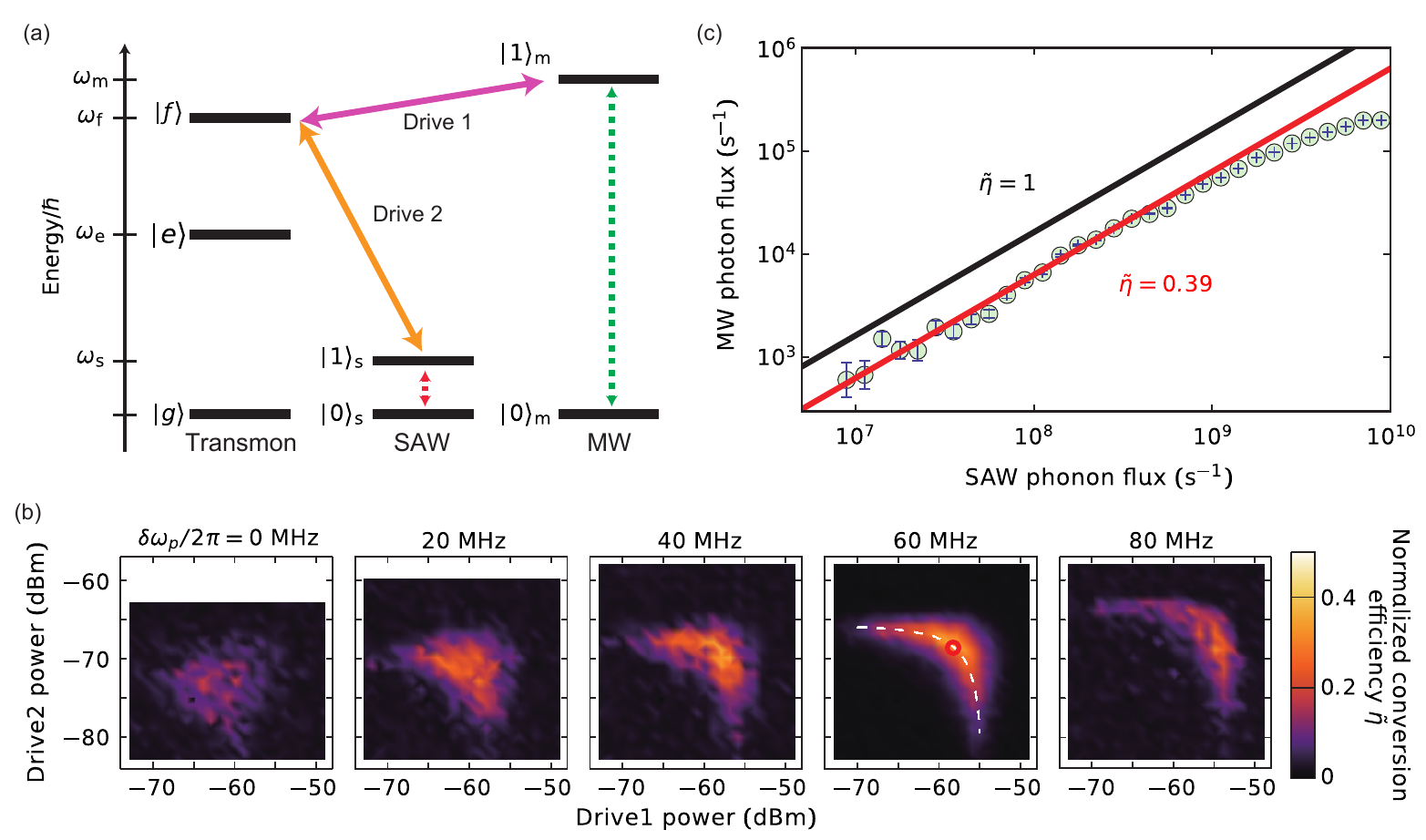}
\caption{Parametrically induced coupling and conversion. (a)~Energy diagram of the hybrid system.
(b)~Optimization of the parametric conversion efficiency $\tilde{\eta}$ as a function of the qubit drive powers applied at port~2.
The drive frequencies $\omega _{d1}$ and $\omega _{d2}$ are varied between panels:
The single photon detuning  $\delta \omega_p \equiv (\omega_{d1}-\omega_{m} + \omega_{f})/2\pi$ are set to {0, 20, 40, 60, 80}~MHz from left to right, while the two-photon detuning $ \omega _{s}+\omega _{d1}+\omega _{d2}-\omega _{m}$ is kept at 0.
The white dashed line in the second panel from the right is a contour on which the AC Stark shift of the qubit $|f\rangle$ state is constant. 
The red circle indicates the optimized condition used in~(c).
(c)~Conversion from the surface acoustic waves to the microwaves~(blue circles with an error bar).
The horizontal axis expresses the input phonon flux to the SAW resonator, and the vertical axis represents the output photon flux from the MW resonator.
The black line depicts the maximum conversion efficiency~($\tilde{\eta}=1$) attainable with the external couplings of the resonators in the present device.
The red line is a linear fit to the data.
The saturation effect observed at high phonon flux is due to the shift of the SAW resonator frequency~[See Fig.~1(d)].
}
\label{fig3}
\end{figure*}

The MW resonator and the SAW resonator are not directly coupled to each other. Nevertheless, we can parametrically induce the coupling by driving the qubit, i.e., the nonlinear element in the system.
The lowest-order parametric coupling is achieved by using the second excited state of the qubit.
We strongly drive the qubit at the frequency $\omega_m - \omega_f$ as illustrated in Fig.~3(a).
This drive~(drive~1) induces the parametric coupling between the MW resonator and the second excited state of the qubit,
\begin{equation}
\hat{U}_{p,m}=\hbar g_{p,m}(\hat{a} \, \hat{\sigma}_{fg} +\hat{a}^\dagger \hat{\sigma}_{gf}),
\end{equation}
where $g_{p.m}$ is the coupling strength.
Similarly, we also drive the qubit at $\omega _{f} -\omega_{s}$.
This tone (drive~2) induces the parametric coupling between the SAW resonator and the qubit
\begin{equation}
\hat{U}_{p,s}=\hbar g_{p,s}(\hat{c} \, \hat{\sigma}_{fg} +\hat{c}^\dagger \hat{\sigma}_{gf}),
\end{equation}
where $g_{p,s}$ is the coupling strength.
Each coupling strength is written as~\cite{Wallraff2014}
\begin{equation}
g_{p,i} = \frac{1}{\sqrt{2}}\frac{g_{i}\alpha}{(\omega _{e}-\omega _{i}) (\omega _{e}+\alpha -\omega _{i})}\Omega _{i},
\end{equation}
where $i = m, s$ and $\Omega _{m}$ ($\Omega _{s}$) is the amplitude of drive~1~(drive~2).
Note that this perturbative treatment is valid for the weak drive strength, where $\Omega _{i}$ is smaller than $| \omega _{e}-\omega _{i}|$ and $|\omega _{f}-\omega _{i}|$.

With the simultaneous drives of the two parametric processes, conversion between the MW photons and the SAW phonons is realized.
Figure~3(b) shows the conversion efficiency from the SAW phonons to the MW photons  as a function of the powers of drives~1 and~2.
The color in the 2D plot indicates the normalized conversion efficiency, $\tilde{\eta}= \eta/\eta_m \eta_s$, where $\eta_m=\gamma_\mathrm{ex}/\gamma =0.20$ and $\eta_s=\Gamma_\mathrm{ex}/\Gamma = 1.6\times 10^{-3}$ are the external coupling factors of the MW and SAW resonators, respectively.
In the series of measurements shown in the panels of  Fig.~3(b), we optimize the conversion efficiency by varying the frequencies of the two drive tones, while keeping their sum fixed.
Naively speaking, one would expect a better conversion efficiency with a higher driving power.
However, the AC Stark effects induced by the drives shift the energy level of the second excited state of the qubit, and thus the detunings of the parametric drives depend on the power of the drives themselves.
In the second panel from the right, we obtain the highest efficiency at the drive condition marked by the red circle. 
The white-dashed line delineates the condition where the sum of the Stark shifts is constant.
The observed `arc' is well fitted by the line, indicating that the large conversion efficiency is obtained when each parametric drive is in resonance with the relevant transition.

The conversion efficiency in the SAW-qubit-MW hybrid system reads~\cite{Painter2012}
\begin{equation}
\eta = \eta _{m}\eta_{s} \frac{4C_{m}C_{s}}{(1+C_{m}+C_{s})^2},
\label{eq00}
\end{equation}
where $C_{m}=4g_{p,m}^2/\gamma\kappa _f$ and
$C_{s}=4g_{p,s}^2/\Gamma\kappa _f$ are the cooperativities of the MW-qubit and the SAW-qubit coupled systems, respectively.
Here $\kappa_f/2\pi = 18$~MHz is the decay rate of the second excited state $|f\rangle$ of the qubit.
In Fig.~3(c), we plot the output MW photon flux $P_m$ from the MW resonator vs.\ the input SAW phonon flux $P_s$. 
The data is taken under the optimized drive power condition indicated by the red circle in Fig.~3(b). 
The photon/phonon fluxes are calibrated based on the Stark shift measurements in the resonators.
The black line depicts the relation $P_m=4\eta _{s}\eta_{m} P_{s}$, which indicates the upper limit of the conversion efficiency with the given external couplings.
The red line is the linear fit to the data, showing the maximum conversion efficiency of $0.39 \times \eta_m \eta_s$.
From the symmetry between drives~1 and 2 in the power dependence of the conversion efficiency observed in Fig.~2(b), as well as from the functional form of Eq.~(\ref{eq00}),  we assume that the two cooperativities are equal at the optimal point. Then, the cooperativities are evaluated to be $C_m=C_s=0.82$, which correspond to the parametrically induced coupling strengths of $g_{p,m}/2\pi = 1.6 \mathrm{~MHz}$ and $g_{p,s}/2\pi = 0.37 \mathrm{~MHz}$ for the MW and SAW resonators, respectively.
The corresponding amplitudes of the drives are $\Omega _{m}/2\pi =1.2 \mathrm{~GHz}$ and $\Omega _{s}/2\pi =1.0\mathrm{~GHz}$, which satisfy the perturbative condition.
The conversion efficiency is currently limited by the smallness of the external coupling of the SAW resonator $\eta_s$.
It can be overcome by using a high-Q SAW resonator or making the IDT electrodes large to enhance the external coupling of the SAW resonator.

\begin{figure}[bt]
   \includegraphics[width=8cm,angle=0]{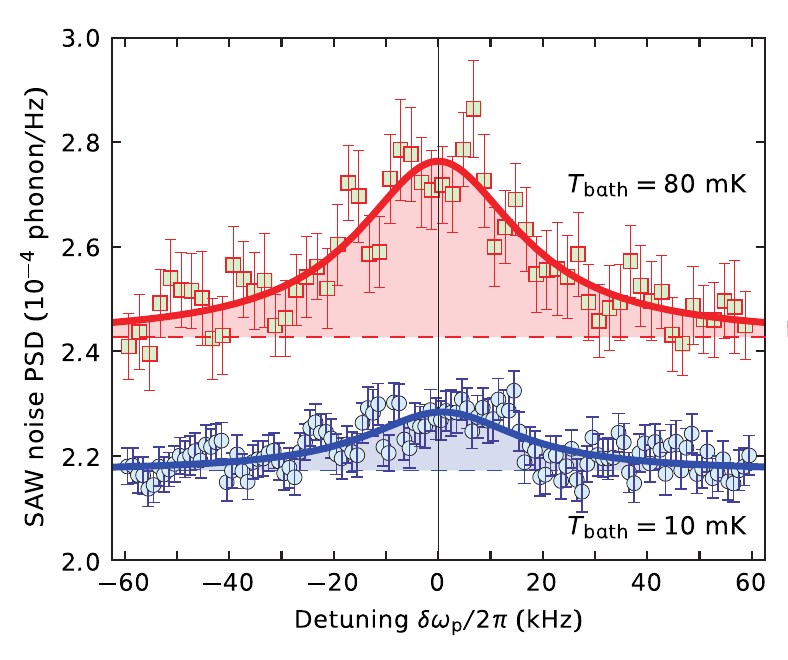}
\caption{Thermal noise power spectral density~(PSD) in the SAW resonator as a function of the parametric-drive detuning  $\delta \omega_{p}=\omega _{s}+\omega_{d1}+\omega_{d2}-\omega _{m}$. Here, we sweep $\omega_{d2}$, while fixing $\omega_{d1}$.
The SAW phonon noise is observed via the parametric up-conversion into the MW photon noise and plotted in the unit of phonon number fluctuation.
The blue circles and the red squares represent the data taken at the nominal bath temperatures $T_\mathrm{bath}$ of 10~mK and 80~mK, respectively.
The solid curves are the Lorentzian fits.
The backgrounds indicated by the dashed lines are the noise floors in the microwave measurement.
The apparent difference in the background level is due to the temperature dependence of the conversion efficiency.
}
\label{fig4}
\end{figure}

Finally we use the parametric conversion to analyze the thermal noise in the SAW resonator near the quantum ground state.
Without any drive on the SAW resonator, the mean phonon number of the thermal excitation of the SAW resonator at the dilution fridge temperature of 10~mK is expected to be $10^{-3}$.
We up-convert the SAW thermal fluctuation to that in the MW resonator through the parametric drives and measure the noise spectrum of the microwave output field using a cascade of a Josephson parametric amplifier~\cite{Yamamoto_JPA} and a traveling wave parametric amplifier~\cite{TWPA}~(See~\cite{supple} for the details).
Figure~4 shows the noise spectral density of the microwave resonator.
The thermal noise in the SAW resonator manifests as an additional noise peak on top of the background.
The linewidth of the peak agrees with $\Gamma$. 
From the peak area, the mean phonon number $\langle n_s \rangle$ in the SAW resonator is evaluated to be $0.57\pm 0.07$, corresponding to the effective temperature of $37\pm 3$~mK, that is significantly higher than the bath temperature of 10~mK, presumably due to the bad thermalization. 
At a higher bath temperature of 80~mK, the mean phonon number increases to $1.8 \pm 0.2$, corresponding to the effective temperature of $85 \pm 6$~mK.
The displacement sensitivity of the measurement is evaluated as 0.2$\mathrm{~am/\sqrt{Hz}}$, two orders of magnitude improvement from the previous report~\cite{delsing2012}.

In summary, we constructed a hybrid quantum system consisting of a SAW resonator, a transmon qubit, and a MW resonator. The interaction between the SAW and MW resonators were mediated by the parametrically driven qubit. We up-converted the SAW phonons to the MW photons and detected the thermal fluctuations of the SAW resonator near the quantum limit. It demonstrated an application of the hybrid quantum system to the ultra-sensitive measurement of the low-frequency SAW signals.

The conversion efficiency may be significantly improved to an order of unity. The external couplings of the resonators can be readily increased, and the SAW resonator quality factor can be at least an order of magnitude better with an optimized design. Moreover, the limitation in the parametrically induced coupling strengths due to the saturation effect of the transmon qubit can be mitigated with other types of nonlinear superconducting circuits.

A hybrid system in the strong coupling regime may also be achieved. 
The resonance frequency of the SAW resonator can be increased closer to the qubit frequency with shorter-period Bragg grating mirrors and IDT electrodes or with other materials which have a higher sound-velocity. 
The coupling strength between the SAW resonator and the qubit can be kept larger than their decay rates. 

The SAW resonator can also be coupled to an optical system via the opto-elastic interaction~\cite{shumeiko2016,Srinivasan2016,cleland2016,Okada}.
The hybrid system of the SAW resonator and the superconducting qubit opens the possibility of an optical access of superconducting qubit.

The authors acknowledge K. Kusuyama for the help in sample fabrication and W. Oliver for providing the Josephson traveling-wave parametric amplifier. 
This work was partly supported by JSPS KAKENHI (Grant Number 26220601), JST PRESTO (Grant Number JPMJPR1429), and JST ERATO (Grant Number JPMJER1601).

\newpage
\clearpage

\onecolumngrid
\appendix
\begin{center}
{\Large\bf{Supplementary Information for \\ 
\vspace{2mm}
Qubit-assisted transduction for a detection of surface acoustic waves \\
\vspace{2mm}
near the quantum limit}}
\end{center}

\renewcommand{\thefigure}{S\arabic{figure}}
\renewcommand{\thetable}{S\arabic{table}}
\renewcommand{\theequation}{S\arabic{equation}}
\setcounter{figure}{0}
\setcounter{equation}{0}

\begin{figure*}[b]
   \includegraphics[width=16cm,angle=0]{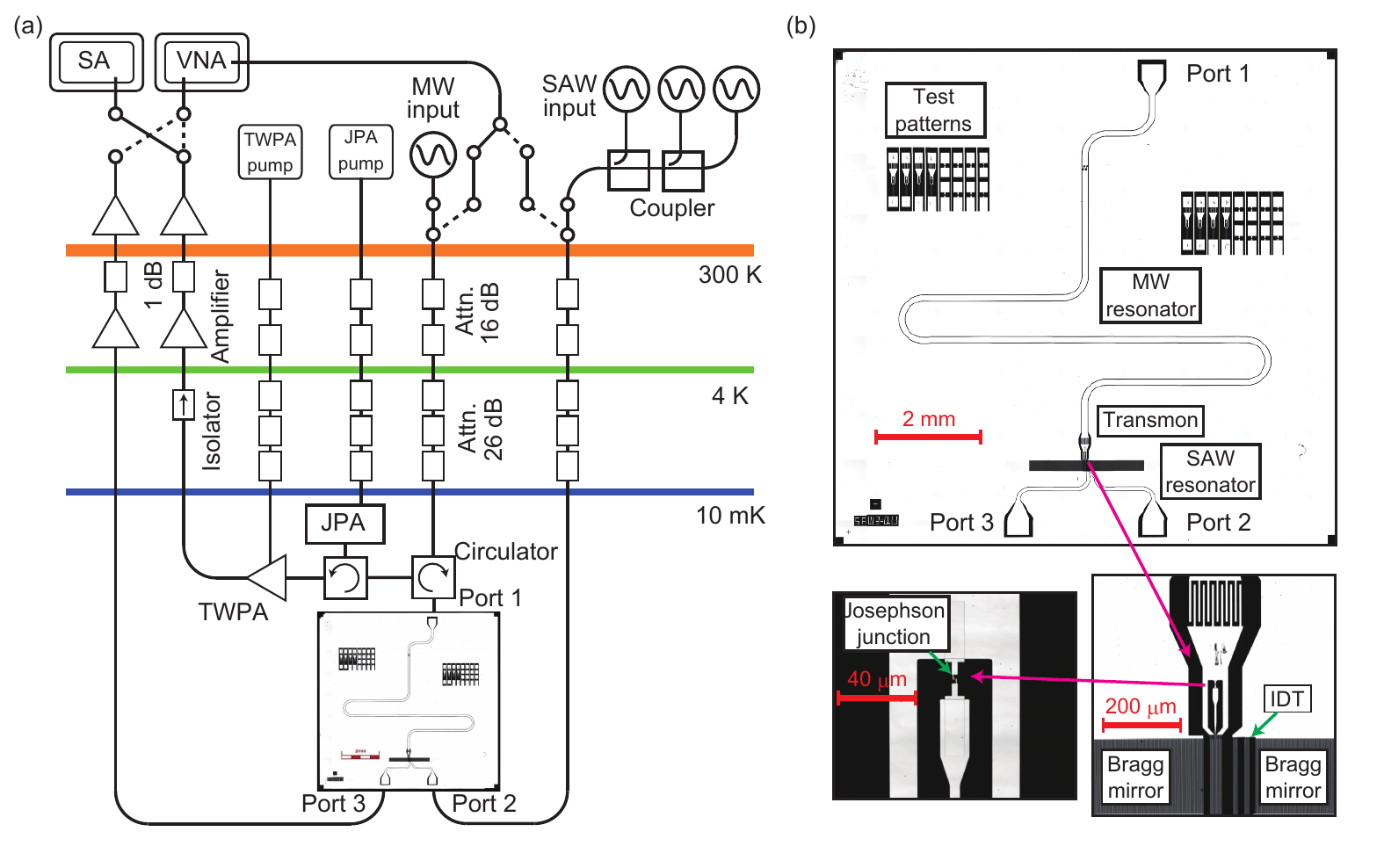}
\caption{
Experimental setups. 
(a)~Wiring for the measurements.
(b)~Photograph of the chip. Insets are the magnifications of the parts.
Note that the IDTs for ports~2 and~3 are on the same side of the SAW resonator, in contrast to the illustration in Fig.~1(a) of the main text.
}
\label{sup6}
\end{figure*}

\section*{Measurement setup}
Figure~\ref{sup6} illustrates the microwave measurement setup for the SAW-MW parametric conversion and the thermal noise detection.
The sample is cooled down to 10~mK in a dilution refrigerator.
Attenuators are mounted on the input coaxial cables at each plate of the fridge to prevent the thermal noise from entering the resonators.  
The qubit drive and the SAW drive are combined in a series of directional couplers and applied via the SAW input line~(port~2).
A probe field for the MW resonator is supplied through the MW input line~(port~1). 
The reflection signal from the MW resonator is observed with a spectrum analyzer (SA) or a vector network analyzer (VNA) through a cascade of a flux-driven Josephson parametric amplifier (JPA)~\cite{Yamamoto_JPA}, a traveling wave parametric amplifier (TWPA)~\cite{TWPA} at base temperature and low-noise HEMT amplifiers at 4~K and room temperature. Port~3 is for the transmission measurement of the SAW resonator.

\section*{Sample}
The circuit is fabricated on a 300-$\mu$m-thick ST-X cut quartz substrate.
The Bragg mirrors and the interdigitated transducers~(IDTs) for the SAW resonator, the coplanar waveguide for the MW resonator, and the capacitance of the qubit are made from a 300-nm-thick evaporated aluminum film. 
They are simultaneously patterned in a wet-etching process.
The Bragg mirrors have 500 fingers each.
The IDTs for the external coupling have a pair of three fingers each, and the IDT connected to the qubit has a pair of eight fingers.
All those fingers have the line width and the spacing of 1~$\mu$m.
The length of the SAW resonator ~(the inner distance between the Bragg mirrors) is 140~$\mathrm{\mu m}$.
The widths of the Bragg mirrors and the IDTs are 200~$\mu$m.

The Josephson junction of the qubit is made from an Al/AlO$_x$/Al junction, which is fabricated by the shadow evaporation technique.
The size of the junction is $140~\mathrm{nm}\times 140~\mathrm{nm}$, and the resistance of the junction at the room temperature is 30~k$\Omega$.
The single-electron charging energy $E_\mathrm{C}$ and the Josephson energy $E_\mathrm{J}$ of the transmon qubit are estimated to be $h \times 230$~$\mathrm{MHz}$ and $h\times 4100$~$\mathrm{MHz}$, respectively.

\begin{figure}[bt]
\includegraphics[width=8cm,angle=0]{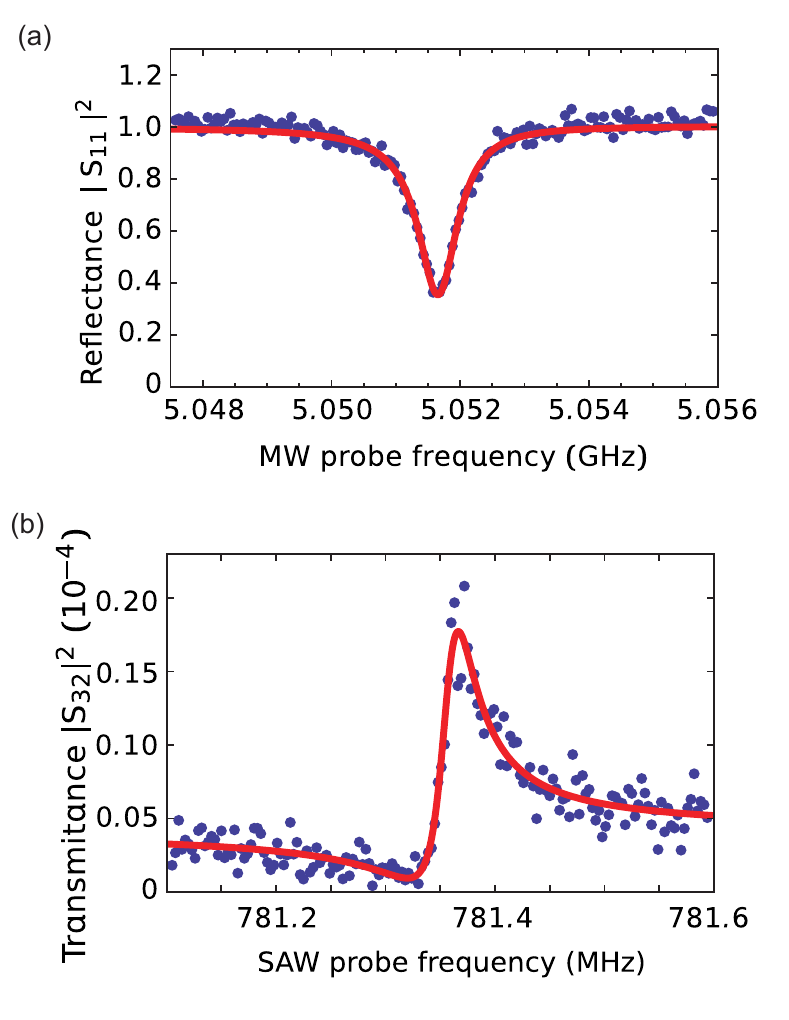}
\caption{
Resonator spectra in the weak probe-power limit.
(a)~Reflection spectrum of the microwave resonator.
(b)~Transmission spectrum of the SAW resonator. 
The transmittance between two IDTs is measured. 
The small transmittance is due to the weak external couplings of the IDTs to the resonator. 
The Fano shape is due to the direct coupling between the two IDTs.
}
\label{sup7}
\end{figure}

\section*{Characterization of resonators}
Figure \ref{sup7} shows the reflection and transmission spectra of the MW and SAW resonators, respectively.
From the results we deduce the frequencies and the linewidths of the resonators presented in the main text.

\begin{figure}[bt]
   \includegraphics[width=8cm,angle=0]{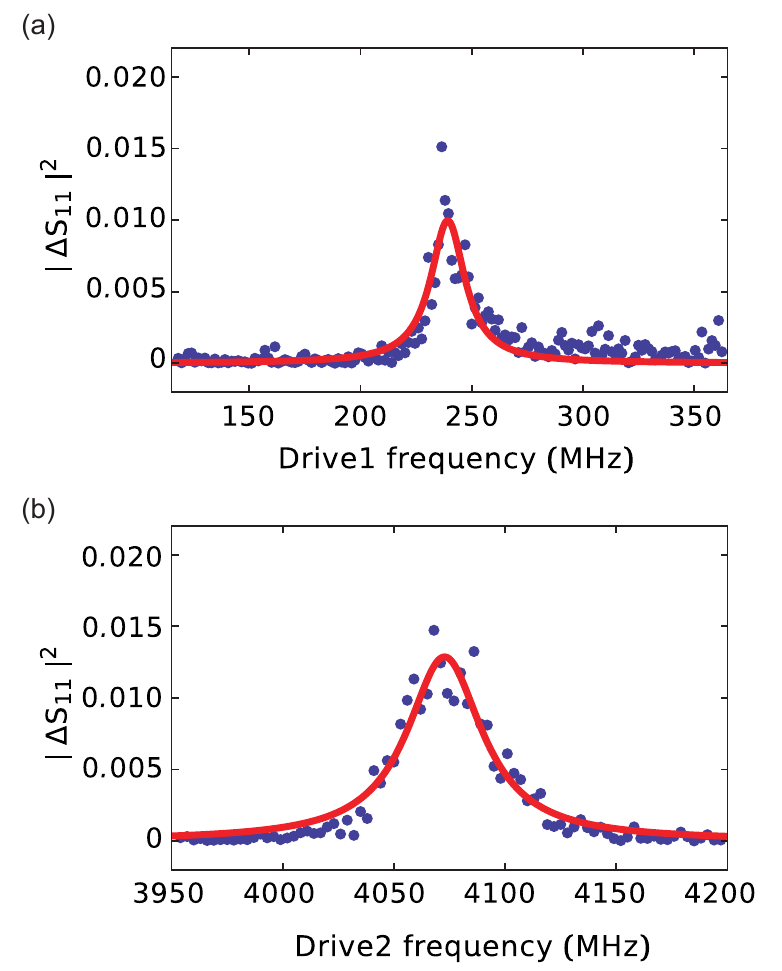}
\caption{Parametric transition spectra in the hybrid system. 
Squared change of the reflection coefficient of the MW resonator is probed at 5.052~GHz as a function of the drive frequency applied to port~2. 
(a)~Parametric resonance between the MW resonator and the qubit.
(b)~Parametric resonance between the SAW resonator and the qubit.
The red curves show the Lorentzian fits.
The resonance frequencies of the parametric coupling are shifted by the AC Stark shift induced by the drives themselves.
}
\label{sup2}
\end{figure}

\section*{Parametric transition spectroscopy}
When we apply a drive field at a frequency close to $|\omega _{f} -\omega _{i}|$, the parametric coupling occurs between $|f, n_i\rangle \leftrightarrow |g, n_i+1\rangle$, where $\omega _{f}$ is the eigenfrequency of the second excited state of the qubit and $\omega_i$~($i = m, s)$ is the resonator frequency.
With the parametric drive at resonance, the resonator is dressed by the second excited state of the qubit, and the reflection coefficient of the resonator changes.
The results in Fig.~\ref{sup2} determine the drive frequencies for the parametric couplings of the qubit with the SAW/MW resonators.

\begin{figure}[bt]
   \includegraphics[width=8cm,angle=0]{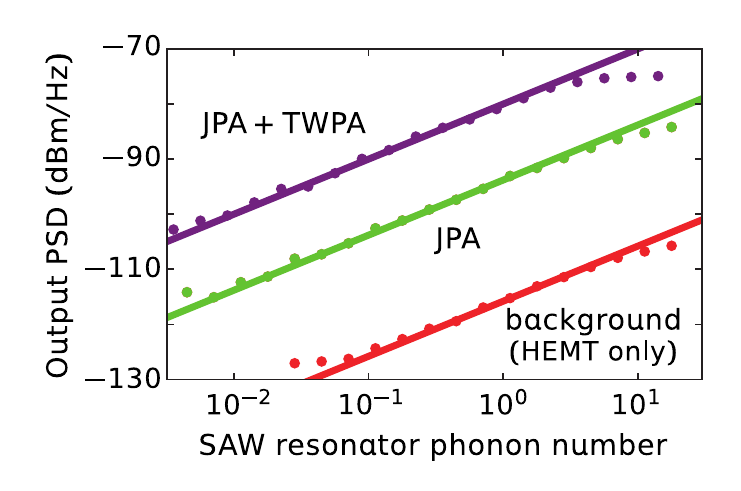}
\caption{Gain calibrations of the JPA and the TWPA.
The lines are the linear fits to the data.
}
\label{sup3}
\end{figure}

\section*{Amplifiers}
Figure~\ref{sup3} shows the gain calibration of the amplifiers.
The horizontal axis is the intra-resonator mean phonon number of the SAW resonator, which is calibrated by the Stark-shift measurement in Fig.~2 of the main text.
The vertical axis represents the power spectral density~(PSD) of the output microwave field at the spectrum analyzer~(Fig.~\ref{sup6}).
The purple, green and red dots show the results with JPA+TWPA, JPA, and without them, respectively.
From the data, the gains of the JPA and the TWPA are determined to be 22~dB and 14~dB, respectively. 
The measurement in Fig.~3 is conducted with JPA+TWPA in the linear region below the saturation point around three intra-resonator SAW phonons.

\end{document}